\documentstyle[psfig,onecolumn]{mn}

\def\be{\begin{equation}}
\def\ee{\end{equation}}

\def\msun{M_{\odot}}

\title[Luminous hot accretion disks]
      {Luminous hot accretion disks}
\author[Feng Yuan]
{\parbox[]{6.in} {Feng Yuan$^{1,2}$\thanks{present address: Max-Planck-Institut
f\"{u}r Radioastronomie, Auf dem H\"{u}gel 69, D-53121 Bonn, Germany. Email:
fyuan@mpifr-bonn.mpg.de}\\ 
\footnotesize
1. Department of Astronomy, Nanjing University, Nanjing 210093, China;
Email: fyuan@nju.edu.cn\\
2. CAS-PKU Joint Beijing Astrophysical Centre, Beijing 100871, China\\}}

\date{Accepted .
      Received ;
     in original form }

\pagerange{\pageref{firstpage}-- \pageref{lastpage}}
\pubyear{2000}

\begin{document}

\maketitle

\label{firstpage}

\begin{abstract}
We find a new two-temperature hot branch of equilibrium solutions for stationary
accretion disks around black holes. In units of Eddington accretion rate 
defined as $10L_{\rm Edd}/c^2$,
the accretion rates to which these solutions correspond are within the range 
$\dot{m}_1 \la \dot{m} \la 1$, here $\dot{m}_1$ is the critical rate of advection-dominated accretion flow (ADAF).
In these solutions, the energy loss rate of the ions by
Coulomb energy transfer between the ions and electrons
is larger than the viscously heating rate and it is the advective
heating together with the viscous dissipation that balances the Coulomb
cooling of ions.
When $\dot{m}_1 \la \dot{m} \la \dot{m}_2$, 
where $\dot{m}_2 \sim 5 \dot{m}_1 < 1$,
the accretion flow remains hot throughout the disk. 
When $\dot{m}_2 \la \dot{m}
\la 1$, Coulomb interaction will cool the inner region of the disk within 
a certain radius ($r_{\rm tr} \sim$ several -- tens of Schwarzschild radii  
or larger depending on the accretion rate 
and the outer boundary condition) and
the disk will collapse onto the equatorial plane and form an 
optically thick cold annulus.
Compared to ADAF, these hot solutions are much more
luminous because of the high accretion rate and
efficiency, therefore, we name them luminous hot accretion disks.
\end{abstract}

\begin{keywords}
accretion, accretion disks -- black hole physics --
galaxies: active  --  galaxies: nuclei
hydrodynamics  -- radiation mechanisms: thermal 
\end{keywords}

\section{Introduction}
There are currently four black hole accretion disk models. They are
the standard thin disk (Shakura \& Sunyayev 1973), slim disk 
(Abramowicz et al. 1988), cooling-dominated hot disk (Shapiro, Lightman, \&
Eardley 1976, hereafter SLE), and 
advection-dominated accretion flow (ADAF) (see reviews by
Narayan, Mahadevan, \& Quataert 1998 and Kato, Fukue \& Mineshige 1998).
Among them, the former two models
are cold, while SLE is thermally unstable. Therefore, ADAF is the only
viable hot accretion disk model up to date.
The success of this model
is in virtue of the two-temperature plasma concept first put forward
by SLE and the realization
to the importance of the advection term in the energy equation of ions
(Ichimura 1977; Rees et al. 1982; Narayan \& Yi
1994, 1995; Abramowicz et al. 1995).
In a typical two-temperature 
ADAF, the accretion rate is 
very low and the Coulomb energy transfer from ions to 
electrons is very inefficient
therefore almost all of the viscously
dissipated energy is stored in the ions and advected into the centre black hole
rather than transferred to the electrons and radiated away.
In other words, in the energy equation of ions,
\be
\rho v T_i \frac{ds}{dr}
=\rho v \frac{d \epsilon_i}{dr}-q^c \equiv q_{\rm adv}=q^+-q_{ie},
\ee
where $s$ and $\epsilon_i$ are the entropy and internal energy of the ions per
unit mass of plasma, $q^c$ and $q^{ie}$ are the compressive heating
and Coulomb energy transfer from ions to electrons, and $q_{\rm adv}$ denotes
the energy advection, we have for a typical ADAF,
\be
q_{\rm adv} \approx q^+ \gg q_{ie}.
\ee

It is well known that the optically thin two-temperature 
advection-dominated accretion
flow (ADAF) solution exists only for the mass accretion rate less
than a critical value $\dot{m}_1$
(Ichimura 1977; Rees et al. 1982; Narayan \& Yi
1995; Abramowicz et al. 1995; Esin et al. 1996, 1997). 
This is because $q^+ \propto \dot{m}$ while $q_{ie} \propto \dot{m}^2$,
i.e., $q_{\rm ie}$ increases faster than $q^+$ with the 
increasing accretion rate.
When the accretion rate reaches a certain critical value,
the Coulomb coupling
between the ions and electrons becomes so efficient that a large
fraction of the viscously dissipated energy is transferred to the electrons 
and radiated away therefore the accretion flow ceases to be an ADAF. 
The critical accretion rate of ADAF $\dot{m}_1$ 
is determined by the balance between
the viscous heating and the Coulomb energy transfer from the ions to the
electrons
\be
q^+ \approx q_{ie}.
\ee
(Narayan, Mahadevan, \& Quataert 1998). 
Due to its low accretion rate and efficiency, an ADAF is unable to
emit significant radiation, i.e., it is a {\em dim} hot accretion disk.

An interesting question is, what will happen when the mass accretion rate 
increases above $\dot{m}_1$. It is in general assumed that in this case, 
hot solutions don't exist and 
the standard thin disk is the only viable solution.
However, this is not true.
From eq. (1), we know that 
in addition to the viscous dissipation $q^+$, 
the compression work $q^c$ is also a term which can heat the ions.
The plasma can remain hot if
the ion cooling rate, $q_{ie}$, is lower than the  sum of $q^+$ and $q^c$.
Since $q^c \propto \dot{m}$, we expect that there must exist
another critical accretion rate below which the accretion flow can be hot.
This critical accretion rate, $\dot{m}_2$, is determined by
\be
q_{ie} \approx q^c + q^+. 
\ee
The critical rate $\dot{m}_2$ could be obviously larger than $\dot{m}_1$ 
if $q^c$ is in the same order to, or larger than, $q^+$,
as we will show by simple analytical estimate and exact numerical
calculation in the following.
And more importantly, there must 
exist another type of hot accretion solutions in addition to ADAF, 
which corresponds
to accretion rate between $\dot{m}_1$ and $\dot{m}_2$. 
In these solutions, the ions cooling rate is lower than the sum of 
$q^+$ and $q^c$ but higher than the viscous dissipation rate
$q^+$. Obviously, compared to ADAF, this new hot solution 
will be much more luminous.

\section{luminous hot accretion disk}

Before exploring deeply this new solution, we  first
give some simple analytical estimate to the range of accretion 
rate within which this solution exists.  
The viscous dissipation, compression work, and
the Coulomb energy transfer in the ions energy equation 
(1) have the following forms,
\be
q^+=\rho \nu r^2\left(\frac{d \Omega}{dr}\right)^2
=\alpha \rho c_s H r^2\left(\frac{d \Omega}{dr}\right)^2,
\ee
\be
q^c=-\rho v p_i d\left(\frac{1}{\rho}\right)/dr,
\ee
and
\be
q_{ie}=\frac{3}{2}\frac{m_e}{m_i}n_en_i \sigma_T \,c\, {\rm ln} \Lambda
\left(kT_i-kT_e\right)\frac{\left(\frac{2}{\pi}\right)^{1/2}+\left(\theta_e+
\theta_i\right)^{1/2}}{\left(\theta_e+\theta_i \right)^{3/2}}.
\ee
(see Dermer, Liang \& Canfield 1991 for $q_{ie}$). Here the subscribes ``i'' and
``e'' denote quantities for ions and electrons, respectively.
The Coulomb logarithm ${\rm ln} \Lambda \approx 15$ for stellar-mass 
black hole sources and $\approx 20$ for AGNs, 
$\theta_i=kT_i/m_ic^2$ and $\theta_e=kT_e/m_ec^2$.
Following Narayan, Mahadevan, \& Quataert (1998),
using the self-similar scaling law obtained by Narayan \& Yi (1995)
(see also Mahadevan 1997), we obtain,
\be
q^+ \approx  3 \times 10^{20} m^{-2} \dot{m} (r/r_g)^{-4},
\ee
\be
q^c \approx 5 \times 10^{8} m^{-2} \dot{m} (r/r_g)^{-3} T_i,
\ee
and
\be
q^{ie} \approx 1.8 \times 10^8 \alpha^{-2} m^{-2} \dot{m}^2 (r/r_g)^{-3}
(T_i-T_e)\theta_e^{-3/2}.
\ee
All the three quantities above are in cgs units.
Here the accretion rate is in units of
Eddington accretion rate defined as $10 L_{\rm Edd}/c^2$, the black hole
mass is in units of solar mass, $M=m \msun$,
$r_g$ denotes the black hole radius, $r_g=2GM/c^2$.
From eqs.(3)(8) and (10), setting the ions temperature as virial,
$T_i \simeq 2 \times 10^{12} \beta (r/r_g)^{-1}$,
where $1-\beta$ is the ratio of the magnetic pressure to the sum
of the magnetic pressure and the gas pressure, 
and $\theta_e \simeq 1/3$,
we obtain the critical accretion rate for ADAF,
\be
\dot{m}_1 \approx 2 \times 10^{12} \alpha^2 T_i^{-1} \theta_e^{-3/2} \approx 
0.4 \alpha^2.
\ee
We set $\beta =0.5$ throughout the present paper, i.e., 
we assume exact equilibrium between
the magnetic and gas pressure.
From eqs. (4), (8), (9), and (10), we obtain $\dot{m}_2$,
\be
\dot{m}_2 \approx \frac{3 \times 10^{12} (r/r_g)^{-1} + 5 T_i}
{8 T_i} \alpha^2.
\ee
If we set $T_i \simeq 10^{12}(r/r_g)^{-1}$, then $\dot{m}_2 \approx
\alpha^2$. In fact, the temperature of ions will be moderately 
lower than the virial value as $\dot{m} \approx \dot{m}_2$
since according to our definition of $\dot{m}_2$, most of the
compression work and viscous dissipation
are used to compensate the Coulomb energy transfer loss rather than
increase the internal energy of ions. Thus,
the value of $\dot{m}_2$ will be much larger than $\alpha^2$. The exact value of
$\dot{m}_2$ can only be obtained by self-consistently solving
the radiation hydrodynamic accretion 
equations by numerical calculation. 

New hot accretion solutions must exist when the accretion
rate is higher than $\dot{m}_1$ but less than $\dot{m}_2$.
In these solutions, the ion cooling rate $q^{ie}$ is so efficient
due to the high accretion rate that the viscous dissipation alone is
not large enough to
compensate it, i.e., the right-hand side of eq. (1) is negative.
It is the compression work of the accretion flow,
another heating source of plasma, together with $q^+$ plays this role.
In other words, the energy advection now is a heating rather than
a cooling term in the ions energy balance equation.
The entropy of the flows therefore {\em decrease} with the decreasing
radii, similar to the Bondi accretion
and the cooling flows in galaxy clusters.
It is also very similar to the electrons in a typical ADAF with
$\dot{m} \ll \dot{m}_1$.
As pointed out by Nakamura et al. (1997), in that case,
the energy advection
by electrons plays a heating rather than cooling role
that compensates the radiative cooling
of electrons. 

Another effect arisen when we numerically solve the 
accretion equations is that $\dot{m}_2$
is actually a function of radius. We will sign $\dot{m}_2$
in the following as
the critical rate independent of radius below which a hot solution exists
{\em throughout} the disk, from the outer boundary to the horizon.
When $\dot{m}$ is greater than $\dot{m}_2$ but less than $ \sim 1$,
we find by numerical calculation presented in the next section
 that below a certain radius the Coulomb
interaction between the ions and electrons will efficiently
cool the inner part of the accretion flow. As a result,
the hot accretion flows will collapse
onto the equatorial plane, forming an optically thick cold annulus.
The exact value of the transition radius depends on the parameters such as
$\dot{m}$ and the outer boundary condition 
of the accretion flow,
as shown by our numerical calculation below. Typically it equals 
several or several tens of $r_g$.

This result, i.e., when the accretion rate reaches a high value
a transition from an outer hot disk to an inner cold disk will occur,
is first anticipated by Pringle,
Rees, \& Pachocyk (1973) qualitatively and further
developed in more detail
by Begelman, Sikora, \& Rees (1987) in the context of quasi-spherical
accretion onto a black hole where the variation of the transition radius
with the accretion rates is obtained although the detailed dynamically
self-consistent solutions are lacked. Narayan \& Popham (1993) also found
this result in one of their examples of the numerical
solutions of the boundary layer of the standard thin disk.

Some features
of this new hot accretion solution can be expected immediately.
The ions temperature should be high, close to or moderately lower than 
the virial value depending on the mass accretion rate $\dot{m}$. 
The efficiency should be high because not only
the viscous dissipation but also the compression work will be transferred 
to the electrons and radiated away. 
Compared to ADAF, the emergent luminosity of this hot accretion disk will be
much higher because of the high efficiency and
accretion rate. 

\section{Numerical calculation Results}

\begin{figure}
\psfig{file=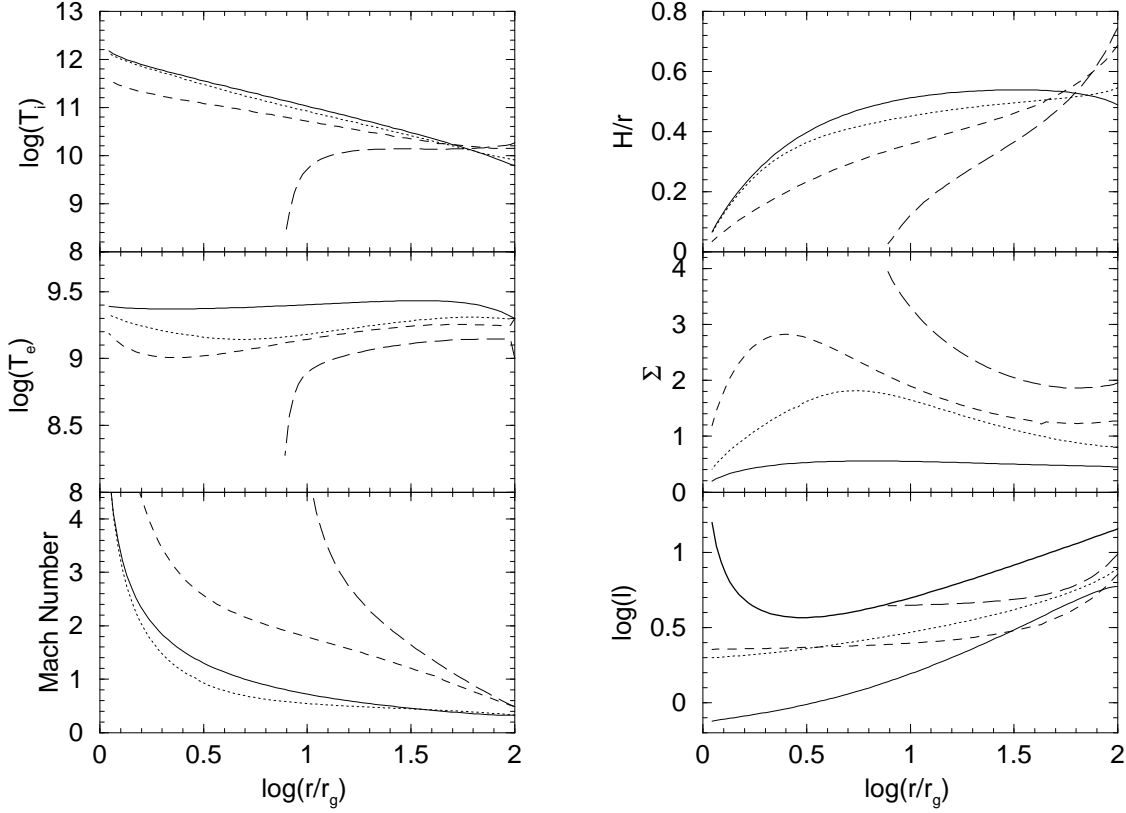,width=0.845\textwidth,angle=270}
\caption{The variations with the radii of Mach number, electrons and ions
temperatures $T_e$ and $T_i$, the specific angular momentum $l$,
the surface density $\Sigma (\equiv \rho H)$, and the ratio of disk
height to radius $H/r$ for some {\em hot}
accretion solutions. Solid (ADAF): $\dot{m}=0.05 < \dot{m}_1,
T_i=6 \times 10^9 K, T_e=2 \times 10^9K, v/c_s=0.4$;
dotted (critical ADAF): $\dot{m}=0.1 \approx \dot{m}_1,
T_i=8 \times 10^9K, T_e=2 \times 10^9K, v/c_s=0.4$;
dashed (new hot solution): $\dot{m}=0.3 < \dot{m}_2, T_i= 1.4 \times 10^{10}K,
T_e=2 \times 10^9 K, v/c_s=0.6$;
long-dashed (new hot solution):
$\dot{m}=0.5 > \dot{m}_2, T_i=1.8 \times 10^{10}K, T_e=10^9K, v/c_s=
0.6$. All are for
$\alpha=0.3, M=10 \msun$ and $r_{\rm out}=100r_g$.
The units of $\Sigma$ and
$T$ are ${\rm g\,cm^{-2}}$ and ${\rm K}$ while
$l$ is in c=G=M=1 units.}
\end{figure}

In this section, we present our exact numerical calculation 
results by self-consistently solving the radiation hydrodynamic
accretion equations.
Paczy\'nski \& Witta (Paczy\'nski \& Witta 1980) 
potential is adopted to mimic the
geometry of a Schwarzschild black hole. 
Steady axisymmetric and two-temperature assumptions to the accretion flow
are adopted.
A randomly oriented magnetic field is assumed to exist in the
accretion flow and the magnetic pressure is in
exact equilibrium to the gas pressure ($\beta=0.5$). 
In this case, the radiation is
very efficient thus $\theta_e \ll 1$ for  $\dot{m} > \dot{m}_1$.
Since thermal pair 
production is very sensitive to the electron temperature,
\be
\dot{n}_{\rm cre} \propto exp(-2/\theta_e),
\ee
we neglect pair effect in the present paper.
The radiation pressure $p_{rad}$ is also neglected because we find
it is less than 10\% of the gas pressure even though when $\dot{m} \sim 1$.
However, $p_{rad}$ becomes comparable to the gas pressure in the transition
region from the hot disk to the thin annulus when $\dot{m} \ga \dot{m}_2$. 

The estimates to $\dot{m}_1$ and $\dot{m}_2$ in section 2
are for a diffusion-type
viscous description where the kinetic viscous coefficient $\nu$ have the
form $\nu = \alpha c_s H$. Another type of 
viscous description widely used in
the literature is assuming that the viscous stress tensor
is proportional to the total pressure,
\be
\tau_{r\varphi}=\alpha p=\alpha (p_{gas}+p_{mag}).
\ee
We will adopt this type of viscous description because in this case
the no-torque boundary condition required in the horizon of the black hole
to solve the set of accretion differential equations can be
automatically satisfied hence significantly simplify our numerical
calculation (Abramowicz et al. 1988; Narayan, Kato, \& Honma 1997).

The equations of mass conservation and the 
hydrodynamic balance in the vertical direction of the disk are,
\be
-4\pi r H\rho v=\dot{M},\hspace{4mm}{\rm and} \hspace{3mm}
 H=c_s/\Omega_{\rm k}
\equiv \sqrt{p/\rho}/\Omega_{\rm k}.
\ee
The radial and axial momentum equations are,
\be
v \frac{dv}{dr}=-\Omega_{\rm k}^2 r+\Omega^2 r-\frac{1}{\rho}\frac
{dp}{dr},
\ee
\be
v(\Omega r^2-j)=\alpha r \frac{p}{\rho}.
\ee
The energy equation for electrons is,
\be
\rho v \left(\frac{d \epsilon_e}{dr}+p_e \frac{d}{dr}
 \left( \frac{1}{\rho}\right) \right)=q_{ie}-q^-,
\ee
where $\varepsilon_e$ denotes the internal energy of the
electron per unit mass of the gas.

The radiation mechanisms we consider include
bremsstrahlung, synchrotron radiation and Comptonization of
soft photons. Assuming the disk is isothermal
in the vertical direction, the spectrum of unscattered photons at a given
radius is calculated by solving the radiative
transfer equation in the vertical direction of the disk based upon the
two-stream approximation (Rybicki \& Lightman 1979).
The result is (Manmoto, Mineshige \& Kusunose 1997):
\be
 F_{\nu}=\frac{2 \pi}{\sqrt{3}}B_{\nu}[1-{\rm exp}(-2\sqrt{3} \tau^*_{\nu}],
\ee
where $\tau^*_{\nu} \equiv (\pi^{1/2}/2) \kappa_{\nu}H$ is the optical
depth for absorption of the accretion flow in the vertical direction with
$\kappa_{\nu}=\chi_{\nu}/(4\pi B_{\nu})$ being the absorption coefficient,
where $\chi_{\nu}=\chi_{\nu, {\rm brems}}+\chi_{\nu,{\rm synch}}$
is the emissivity, and $\chi_{\nu, {\rm brems}}$ and
$\chi_{\nu,{\rm synch}}$ are the bremsstrahlung and synchrotron
emissivities, respectively.  Then the local radiative
cooling rate $q^-$ reads as follows:
\be
q^-=\frac{1}{2H}\int d \nu \eta(\nu)2F_{\nu},
\ee
where $\eta$ is the energy enhancement factor first introduced by Dermer,
Liang \& Canfield (1991) and modified by Esin et al. (1996).

We numerically solve the above coupled radiation hydrodynamic equations (1), 
(5) -- (7), and (15)--(20).
The solutions must satisfy the no-torque condition 
at the horizon, a sonic point condition at a sonic point, and
the outer boundary condition (OBCs) at a certain outer boundary $r_{\rm out}$.
We here choose OBCs as the temperature of ions and electrons, $T_{i,e}$,
and the ratio of the radial
velocity of the flows to the local sound speed, $v/c_s$ at $r_{\rm out}$.
The numerical approach is presented in Yuan et al. (2000) (see also Nakamura
et al. 1997; Matsumoto et al. 1997). We would like to emphasize here that
the OBCs are again found to play an 
important role in determining the dynamics of the solution,
as pointed out by Yuan (1999) in the general context
of optically thin accretion flows.
Both the exact value of $\dot{m}_{1,2}$ and the transition radius
between the outer hot disk and the inner cold thin annulus when $\dot{m}
\ga \dot{m}_2$ depend on OBCs. The effects of OBCs are expected to play a 
more important role in calculating 
the emergent spectrum (Yuan et al. 2000). 
Since we concern here only the most general dynamics,
we don't investigate in detail the effect of OBCs in the present study.

Figure 1 shows the dynamic features of four {\em hot} 
solutions with different accretion rates. 
The parameters are $\alpha=0.3, m=10$. The outer boundary is set at 
$100r_g$, a relatively small radius,
since we first concentrate on the outer cold disk plus
inner hot disk configuration. This two-components configuration
is believed to be the most promising geometry
for the hard state of the Galactic black hole candidates and Seyfert galaxies
(see Zdziarski 2000 for a review). The solid, dotted, dashed, and
long-dashed lines are for $\dot{m}=0.05, 0.1, 0.3$ and $0.5$, corresponding
to a typical ADAF with $\dot{m} < \dot{m}_1$, 
a critical ADAF with $\dot{m} \approx \dot{m}_1$, a new hot
accretion solution with $\dot{m} < \dot{m}_2$, 
and a new hot solution with $\dot{m} > \dot{m}_2$ 
(a transition to an inner cold thin annulus occurs in this case),
respectively. We find from the figure that the ions temperature
decreases with the increasing $\dot{m}$. This is because the two heating terms
$q^c$ and $q^+$ in the ions energy equation 
are both proportional to $\dot{m}$ while
the cooling term $q_{ie} \propto \dot{m}^2$, therefore, a larger
and larger fraction of energy is transferred to the electrons and
radiated away rather than stored in the 
plasma to increase the internal energy of ions.
This is also the reason why the ions temperature
$T_i$ increases slower inward with increasing $\dot{m}$. 
For the long-dashed line which denotes a hot solution with a 
transition to an inner cold annulus,
the temperature of the ions in the hot disk part
almost remains constant.
This shows that the sum of the compression work and the viscous dissipated
energy are nearly equal to the Coulomb energy transfer from ions to electrons.

\begin{figure}
\psfig{file=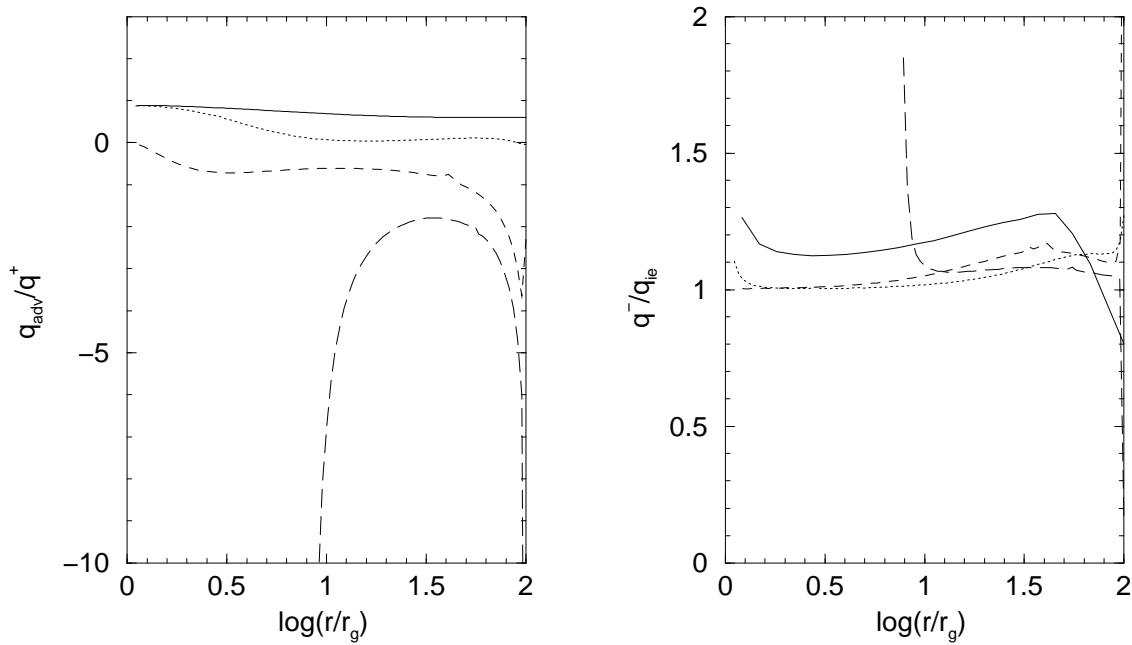,width=0.845\textwidth,angle=270}
\caption{Energy balance relationship for ions(left) and electrons(right)
for the four solutions presented in Figure 1.}
\end{figure}

The two plots in Figure 2 show the corresponding energy
balance relationship for ions (left)
and electrons (right) for the four solutions presented in Figure 1.
The left plot shows the variation of the 
``advection factor'' $f \equiv q_{adv}/q^+$ with radii. 
We note that $f$ is in general set as an ``average'' 
value over radius $r$ in a lot of applications of ADAF
(e.g. Narayan, McClintock, \& Yi 1996; 
Esin, McClintock, \& Narayan 1997; Narayan et al. 1998). 
By self-consistently solving the radiation hydrodynamic
coupled equations, we find $f$ is
typically a sensitive function of radius (see also Figure 5 below) 
if $\dot{m}$ is not very small. 
From Figure 2 we see that 
$ f \approx 1$ throughout the disk for a 
typical ADAF especially with $\dot{m} \ll \dot{m}_1$, 
therefore, the ``average'' advection factor
is a good assumption in this case (Narayan, McClintock, \& Yi 1996; 
Narayan et al. 1998). However,
for the critical ADAF solution, which we define as the solution with 
$f \sim 0$ in ``most'' 
region of the disk, $f$ differ significantly 
from its ``average'' value $\sim 0$
in the innermost region of the 
disk which is the most important part of an ADAF, increasing 
from $\sim 0$ to $\sim 1$ when the flow is accreted inward.
Thus, in this case the assumption of a constant $f$ 
may be a dangerous approximation 
(Esin, McClintock, \& Narayan 1997).
For the two luminous hot accretion solutions with $\dot{m} > \dot{m}_1$,
$f < 0$, as we expected above. 

\begin{figure}
\psfig{file=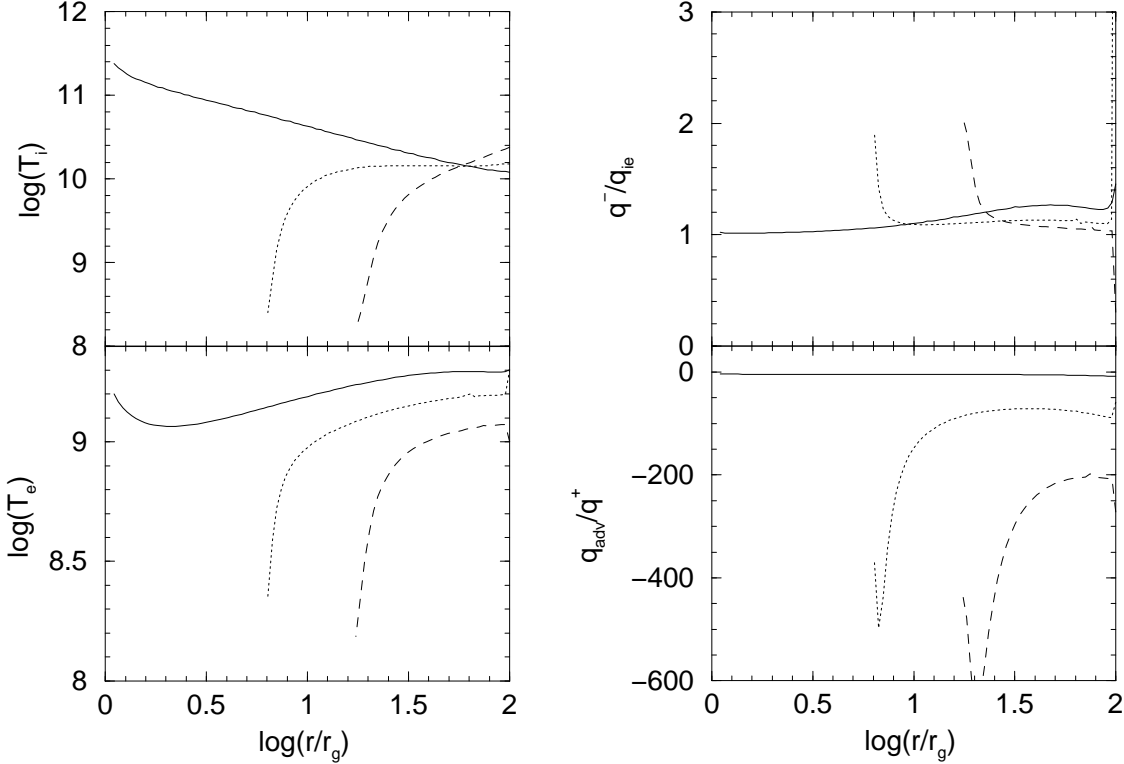,width=0.845\textwidth,angle=270}
\caption{Three more examples of the new hot accretion solutions.
Solid: $\alpha=0.1, \dot{m}=0.3, T_i=10^{10}K, T_e=2 \times 10^9K,
v/c_s=0.8$; dotted: $\alpha=0.01, \dot{m}=0.5, T_i=1.5 \times 10^{10}K,
T_e=2 \times 10^9K, v/c_s=0.8$; dashed: $\alpha=0.01, \dot{m}=1, T_i=2.4
\times 10^{10}K, T_e=10^9K, v/c_s=0.8$.}
\end{figure}

For the energy balance of electrons,
from the right plot of Figure 2, we see that for a typical ADAF with $\dot{m} 
\ll \dot{m}_1$, the radiated energy is larger than the transferred energy from
the ions by Coulomb collision, as first pointed
out by Nakamura et al. (1997). In this case, the electrons must
loss their entropy to balance the radiative loss, like the ions in our new
hot solution. However, when $\dot{m}$ approaches or becomes greater than 
$\dot{m}_1$, we find that $q^- \approx q_{ie}$ approximately holds
(However, see Figure 5 below).

The long-dashed line shows the solution with $\dot{m}>\dot{m}_2$.
In this case, the accretion rate is so large that the hot solution
can only exist beyond a certain radius where the Coulomb energy transfer
is still weaker than the sum of the viscous dissipation and compression
work. Within this transition radius, Coulomb coupling becomes
so efficient that the disk can not remain hot. The only viable solution in
this case is the cold optically thick disk, therefore the disk 
will collapse rapidly onto the equatorial plane and 
form a cold thin annulus. The structure and 
the location of the transition region
are determined by the parameters and the outer boundary conditions
of the flows. Since we assume the gas pressure is much larger than the radiation
pressure in our equations, our integration can't continue after the 
transition
from the hot disk to the cold disk to track the cold disk solution
because the radiation pressure
can not be neglected in that case.
In fact, we find that in the transition region the radiative pressure
begins to become comparable to the gas pressure.

Significant energy will be released in the
transition region, as pointed out by Pringle, Rees, \& Pacholczyk (1973). First,
the magnetic flux tubes will have to expand out of the
plane and strong magnetic reconnection will occur. Second,
in the transition process, the accretion flow cools
rapidly from a temperature close to virial to a state with a much
lower black body temperature, thus the internal energy of the accretion flow
will be released at a luminosity
\be
L \sim \frac{\dot{m}\dot{M}_{\rm Edd}}{m_i}kT_i,
\ee
where $T_i$ is the ions temperature just outside
of the transition radius. Third, the radial velocity of the accretion flow 
must drop down
rapidly in the transition process  
because of the large radiation pressure gradient force
and the centrifugal force.
A shock may happen in the transition region if the radial velocity
of the accretion flow before the transition is supersonic. 
The pre-shock kinetic energy
of the accretion flow will be converted into thermal energy and released.
This part of energy is at least in the same order with
or several times larger 
than $L$ in eq. (21), depending on the value 
of the Mach number of the accretion flow just outside of the transition radius.
 Investigating where these energy go is obviously a very interesting subject.
One possibility should be the formation of a hot corona hovering above 
the inner cold annulus. Different with the hot accretion flow before the
transition, the energy distribution of the electrons
in this corona should be in a power law form at least near the transition radius
due to the magnetic reconnection and shock acceleration,
while the power law energy distribution seems to be a crucial 
factor in modeling 
the high and very high states of black hole X-ray binaries (hereafter 
BHXBs; Zdziarski 2000).

The soft black body photons radiated from the 
inner cold annulus can partially enter 
the outer hot disk to serve as the seed photons of Comptonization. In our
calculation we neglect this cooling effect. 
We expect in this case the transition radius
will move outward slightly but such two-components configuration still holds.
One reason is that from Figure 1 we find 
the scale height of the hot disk just outside of the
transition radius is low thus only a small fraction of black body
photons can be intercepted by the hot disk. 
In addition, the significant energy released in the transition region
will serve as a heating source of the outer hot disk, which would
partially cancel the cooling effects of Comptonization of black body 
soft photons.

\begin{figure}
\psfig{file=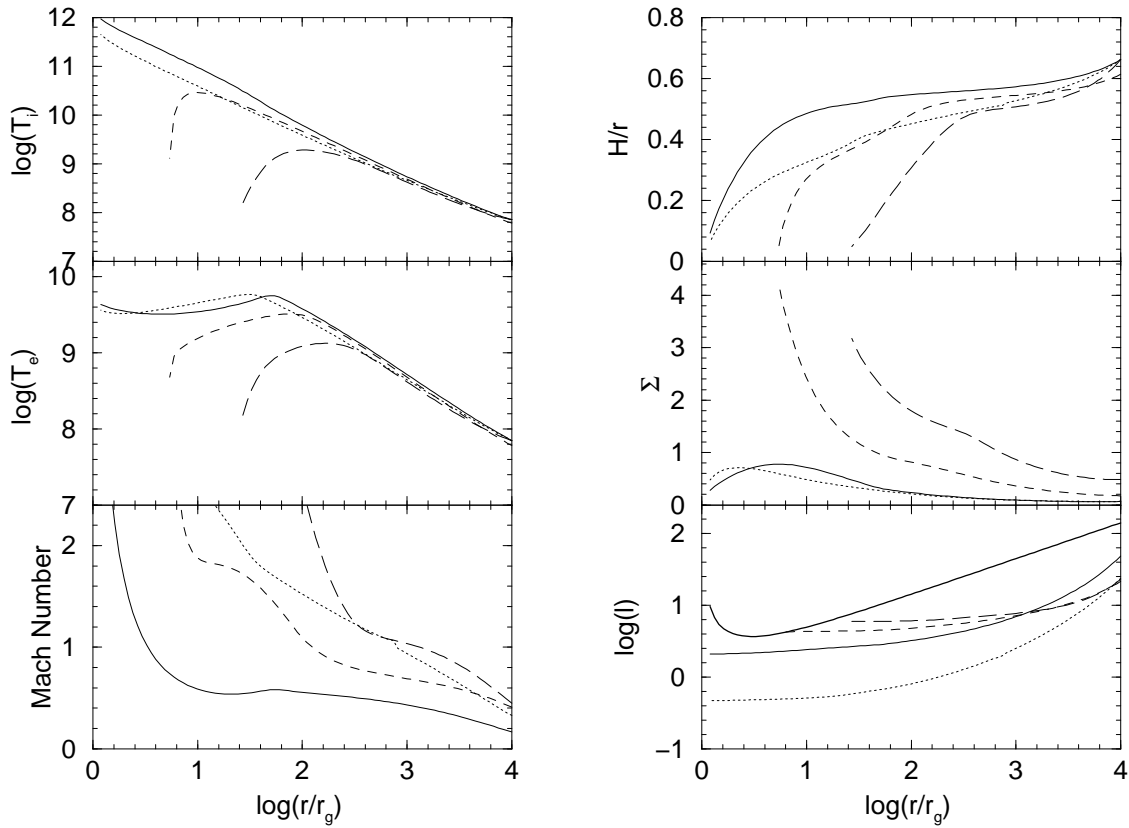,width=0.845\textwidth,angle=270}
\caption{Three hot solutions with $r_{\rm out}=10^4r_g$ and $M=10^8
\msun$. Other parameters are $\alpha=0.1, \beta=0.5$. Solid (critical ADAF):
$\dot{m}=0.05 \sim \dot{m}_1,
T_i \approx T_e \approx 7 \times 10^7{\rm K}, v/c_s=0.2$;
dotted (new hot solution): $\dot{m}=0.1,
T_i \approx T_e \approx 7 \times 10^7{\rm K},
v/c_s=0.5$; dashed (new hot solution):
$ \dot{m}=0.3, T_i \approx T_e \approx
 6 \times 10^7{\rm K}, v/c_s=0.5$; long-dashed (new hot solution):
$T_i \approx T_e \approx 7 \times 10^7 {\rm k}, v/c_s=0.55$.}
\end{figure}

Figure 3 gives some more examples of the luminous hot accretion solutions
with lower viscous parameter $\alpha=0.1$ and 0.01. We see from the figure that 
even when $\alpha$ is as low as $0.01$, the new hot accretion solution
still exists for $\dot{m} = 1$.

All above results are for a black hole with stellar mass and the 
outer boundary is small, $r_{\rm out}=100r_g$.
The results are qualitatively the same for a black hole with galactic mass
and a much larger outer boundary, as shown by Figures 4 and 5 for
the dynamics and energy relationships, respectively. The four solutions 
are for $M=10^8\msun$ and $r_{\rm out}=10^4r_g$.  The solid, dotted, dashed, and
long-dashed lines are for $\dot{m}=0.05, 0.1, 0.3$ and 1. 
They correspond to an ADAF with $\dot{m} \sim \dot{m}_1$,
a luminous hot solution with $\dot{m}_1 < \dot{m} < \dot{m}_2$, 
and two luminous solutions with $\dot{m} > \dot{m}_2$.
Compared to Figures 1 \& 2, we find that $q_{adv}/q^+$ and $q^-/q_{ie}$
are more sensitive to the radius. Especially, a ``peak'' arises in each
of the four lines in the two plots in Figure 5.
This is because the peak in the electron temperature profile leads
to a local minimum in $q_{ie}$.
This result indicates that when the outer boundary is far away from the 
hole, it is hard to make any approximation such as 
$f \equiv q_{adv}/q^+ \approx const.$
or $q^- \approx q_{ie}$. Exact self-consistent global
solutions are needed to calculate the dynamics and the emergent spectrum.

Throughout the present paper, we only take into account bremsstrahlung,
synchrotron, and their Comptonization. For a corona lying above a cold
disk, Comptonization of soft photons from the cold disk should also
be included. But we can't conclude that this cooling 
effect will deduce the maximum
accretion rate below which our luminous hot accretion solution exists
because some extra heating processes in addition to viscous dissipation,
such as magnetic reconnection, would play an opposite role. 

The exact values of $\dot{m}_1$ and $\dot{m}_2$ are hard to determine. 
One reason is because the advection factor $f$ is the function
of radii. More importantly, they are sensitively dependent on
the outer boundary conditions. All the values given below should
only be considered as the approximate values.
Our numerical calculation results show
that when the outer boundary is set at $100r_g$, we have
$\dot{m}_1 \approx 0.25 \alpha^{0.7}$ and 
$\dot{m}_2 \approx 1.5 \alpha^{0.7}$.
The above formula have different dependence on $\alpha$
compared to those obtained in section 2.
One reason is because we set the
outer boundary at a relatively small radius, $100r_g$. As is well known,
the self-similar scaling law $\rho \propto \alpha^{-1}$ fails
and physical quantities such as radial
velocity and density are less sensitive to $\alpha$ in the
inner region of the disk. Another reason is due to 
different viscous descriptions.
In our case, the specific angular momentum of
the accretion flow decreases rapidly away from the outer boundary, but almost
remains constant in the inner region of the disk (see also
Abramowicz et al. 1988; Nakamura et al. 1997). Thus the viscosity
plays a smaller role compared with the case of the diffusion type
viscous description. When the outer boundary is set at $10^4 r_g$,
we find $\dot{m}_1 \sim 0.05, 10^{-2}, 10^{-4}$ for $\alpha=0.1, 0.01, 0.001$
and $\dot{m}_2 \sim 5\dot{m}_1$.
The values of $\dot{m}_1$ are obviously 
different with $\dot{m}_1 \sim \alpha^2$ in Esin, McClintock,
\& Narayan (1997) but are similar with Nakamura et al. (1997). 
The discrepancy is again due to different viscous descriptions. 

\begin{figure}
\psfig{file=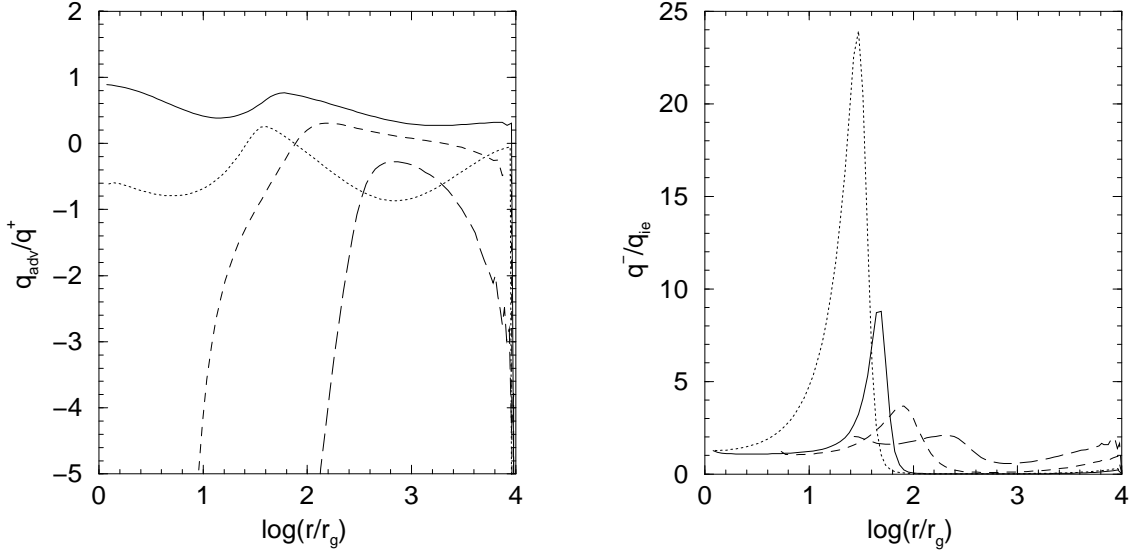,width=0.845\textwidth,angle=270}
\caption{The energy relationship for the four solutions shown in Figure 4.}
\end{figure}

Up to now, we don't concern the stability of the solution.
From the surface density plots in Figures 1 \& 4, we find that the surface 
density of the disk increases with the increase of the accretion rate,
\footnote{In Figure 4 the solid line shows higher 
surface density compared with the dotted line
in some region of the disk although it corresponds to a lower accretion rate. 
This is because these two solutions correspond to different outer
boundary conditions. The specific angular momentum of the flow 
denoted by the solid line at the 
outer boundary $10^4r_g$ is relatively high thus 
the accretion is of disk-like. While for the dotted line, the angular momentum
is relatively low thus the solution is of Bondi-like characterized 
by a much larger sonic radius and lower surface density. See Yuan et al.(2000)
for details.}
therefore the solutions are viscously stable.
The thermal stability analysis is a little difficult. 
Optically-thin hot accretion flow is normally thermally 
unstable when the advection 
in the energy equation is not included in the stability analysis (Piran 1978).
Abramowicz et al (1995) and Narayan \& Yi (1995) show that the optically-thin
hot ADAF is thermally stable because of the inclusion of the energy
advection. Our model is along the line of ADAF, with energy advection is 
explicitly included in the energy equation. So we expect our new
hot accretion solution is also thermally stable.
More convincing thermal stability 
analysis is needed and is a subject of a future investigation. 

In the $\dot{m}$ vs. $\Sigma$ plot, an ``S'' curve is usually 
found for an optically thick
accretion flow, thus for certain choices of $\dot{m}$, the only solution
available is then an unstable one (Abramowicz et al. 1988; 
Chen \& Taam 1993). It is often argued that in this case, 
the accretion flow will be forced into a limit cycle behavior
where the flow oscillates between the two stable solutions. We note that
the ``unstable range'' of $\dot{m}$ is approximately in the range of 
$0.01 \la \dot{m} \la 1$, similar to the range within which our new solution
exists. Therefore, we propose that possibly the accretion flow would
enter into the luminous hot solution rather than enter into the limit cycle.

\section{Summary and discussion}

We find a new branch of hot accretion disk solution. This solution 
corresponds to the accretion rate higher than the critical rate of 
ADAF $\dot{m}_1$ but less than 1 (all the accretion rates are 
in units of Eddington accretion rate defined as
$10L_{\rm Edd}/c^2$).
In these solutions, the viscous dissipated energy is less than the Coulomb
coupling between the ions and electrons and 
it is the advective {\em heating} together with 
viscous dissipation that balances the Coulomb energy transfer 
from ions to electrons. When the accretion rate $\dot{m}$ 
is within the range $\dot{m}_1 \la \dot{m} \la \dot{m}_2 $, where
$\dot{m}_2 \sim 5 \dot{m}_1 < 1$, the hot solution can extend throughout
the disk. When $\dot{m}_2 \la \dot{m} \la 1$, 
within a certain radius of the disk,
the Coulomb cooling of the ions is so strong that even the sum of the viscous
dissipation and compression work can not balance it,
therefore, the hot accretion flow will be cooled by the 
Coulomb cooling and
collapse to form a cold thin annulus.

The new hot solutions are physically quite different with ADAF although
the equations describing them are exactly the same. 
First, they correspond to different ranges of accretion rate.
For a standard ADAF, the viscous dissipation rate is significantly greater
than Coulomb cooling rate, therefore
almost all the viscously dissipated energy is stored in the plasma as
entropy. With the increase of the
accretion rate, more and more viscously dissipated energy is lost
through Coulomb cooling. When the
accretion rate reaches a certain rate, the viscously dissipated energy
is balanced by the Coulomb cooling.
This accretion rate is defined as the critical 
rate of ADAF (Narayan, Mahadevan, \& Quataert 1998). 
In terms of the 
the advection factor $f$, which is defined as the ratio of the energy advection
rate to the viscous dissipation rate, when the accretion rate is 
lower than $\dot{m}_1$, the viscously dissipated
energy is higher than the Coulomb cooling, hence $q_{\rm adv}$ 
and further $f$ are all positive.
This point is justified by the $q_{\rm adv}/q^+$ plots in figures 2, 3, 
and 5 in the present paper as well as in Narayan \& Yi (1995) and Esin, 
McClintock, \& Narayan (1997). When the 
accretion rate is greater than this
critical value, it is widely assumed in all previous literature that any hot
solution can't exist and the only available
solution is the standard thin disk (e.g., Esin, McClintock, \& Narayan 1997). 
However, I find this is not the case.
When the accretion rate is greater
than the critical value of ADAF, hot solutions still exist up to another
(higher) accretion rate.  In
this case the viscous dissipation alone can't balance the Coulomb
cooling,  but the sum of the viscous
dissipation and the compression work can do it. In other words, the new
critical accretion rate is
determined by the balance between the Coulomb cooling and the sum of
viscous dissipation and
compression work.

The difference between ADAF and my solution can also be understood in
the language of entropy. In the 
standard ADAF, the entropy of the accretion flow increases inward while in my
solution the entropy decreases
with the decreasing radii. In the standard ADAF, the energy advection serves
as a cooling term  in the Lagrangian
point of view, while in my solution the energy advection plays a heating
role.  It is the decrease of the entropy of the plasma
that  partially supply the radiation of the accretion flow.  In this
case, my solution is dynamically more similar to the cooling flow in
elliptical galaxies rather than to the ADAF.

Up to date we have four viable accretion disk models. They are the standard thin
disk, slim disk, ADAF, and the luminous hot disk. 
These four models belong to two
series, with the former two models being cold and the latter two
being hot. The transition between the two models 
in each series is due to the variation
of the accretion rate. The transition from the standard thin disk
to the slim disk corresponds to the transition of the advection factor
$f$ from $f \approx 0$ 
to $f > 0$ when the accretion rate passes across $\sim 1$, 
while the transition from ADAF to our
new hot solution corresponds to the transition from $f > 0$ to $f <0$
when the accretion rate passes across
the critical value $\dot{m}_1$
\footnote{So $\dot{m}_1$ should be the accretion rate
at which SLE model is most reasonable applicably since 
$f = 0$ is assumed in SLE.}.
In addition, the cold series exists for almost any value
of accretion rate while the hot series exists only for accretion rate 
approximately less than the Eddington rate $10L_{\rm Edd}/c^2$.
Figure 6 shows the corresponding
accretion rate and efficiency of
these four accretion disk models.
For simplicity we
assume the efficiency of thin disk is $0.1$. 

In previous works (Abramowicz et al.
1995; Chen et al. 1995; Narayan \& Yi 1995; Esin, McClintock, \& Narayan 1997)
both the local and global analysis didn't find the new hot solution. 
This is because they {\em a priori} assumed that the advection factor $f$ must
be positive. They neglected the case where $f$ is negative.

The formation of the luminous hot accretion disk has two possible panels.
If the initial temperature of the accretion gas at the outer
boundary is low, the flow would prefer a cold thin disk first.
In this case, the hot disk could be formed through
the transition from the thin disk.
In the present study we didn't concern the exact physical
mechanism of the transition.
The general proposals
put forward for the thin disk-ADAF transition, such as
evaporation (Meyer \& Meyer-Hofmeister 1994) 
and turbulent diffusive heat transport (Honma 1996),
are still possible mechanisms. In addition, the secular instability
present in the cold disk inner region (Lightman \& Eardley 1974)
could also be a promising mechanism.
As argued by Thorne \& Price (1975), this instability
could swell this optically-thick
radiation-pressure dominated region to a hot, gas-pressure dominated, optically
thin region. Since this instability usually 
appears when the accretion is moderately
high, $\dot{m} \ga 0.1$ for example, this hot optically thin region is 
most possibly described  
by the luminous hot solution since ADAF doesn't exist under such high
accretion rates.

In this context, we would like to note that when the mass accretion rate 
is higher than the critical rate of ADAF $\dot{m}_1$ and when 
the initial temperature of the flow is low at the outer boundary, 
a sandwich model
whereby a corona lies over a cold disk is also viable in addition to
our new hot solution. Then a natural question is which of the two solutions
would be applicable. Our answer is that it is determined by the details
of the mechanism for the formation of the hot solution. If the hot solution is 
due to the evaporation of the cold disk and if the evaporation efficiency is
not high, the sandwich solution will be chosen by the nature. If the 
hot solution is due to the secular instability, or the turbulent diffusive heat 
transport, or the evaporation with high efficiency, our new hot solution should
be chosen. Note in the sandwich solution, there exists the possibility
that the corona might be described by the
luminous hot solution rather than ADAF.

\begin{figure}
\psfig{file=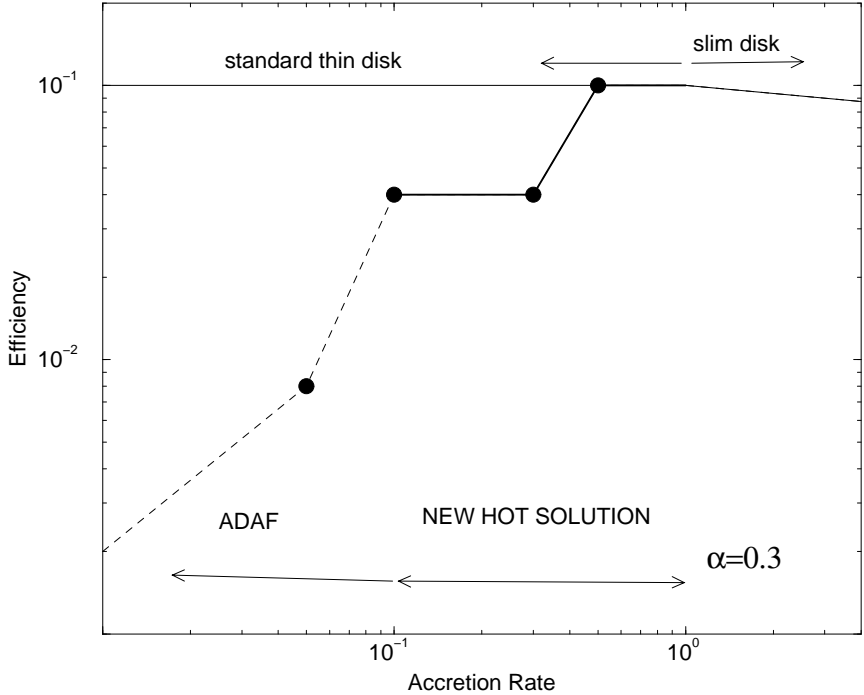,width=0.645\textwidth,angle=270}
\caption{The corresponding accretion rate and efficiency of
four accretion disk models, namely the standard thin disk and slim disk 
(upper thin solid line), ADAF (dashed line) and our new hot 
disk (thick solid line) for viscous parameter $\alpha=0.3$. 
The accretion rate is in units of Eddington
accretion rate defined as $10L_{\rm Edd}/c^2$. 
The four dots in the figure are for the four solutions
presented in Figure 1. Note that if we adopted a smaller 
$\alpha$, the corresponding 
range of accretion rate of the new hot solution would be much larger
than presented in this figure.}
\end{figure}

If the temperature of the accretion matter at the outer boundary 
is high, then the hot accretion solution will form from the beginning
of accretion.
The new hot solutions are possibly very common 
in the centre of galaxies because: 
(1) the temperature of gas there is generally as high as $10^6 \sim 10^8K$; 
(2) if the practical accretion rate is moderately large and 
if the viscous parameter $\alpha$ is not very high, this new hot solution is
the only feasible hot solution since the critical accretion
rate of ADAF, $\dot{m}_1 \sim \alpha^2$ (for diffusion-type viscous 
description), is very small, 
while the new hot solution can exist under 
an accretion rate as high as $\dot{m} \sim 1$ even though 
$\alpha$ is as low as 0.01 (see Fig. 3). 
For example, even for $\alpha=0.1$, an ADAF exists
only when $\dot{m} \la 0.01$, while above this accretion rate,
$0.01 \la \dot{m} \la 1$, the new hot solution holds.

We didn't consider the possible existence of outflow in our model.
Strong outflows are assumed to be present in ADAF due to its positive 
sign of the Bernoulli parameter and the inflow-outflow
solution (ADIOS) was developed by Blandford \& Begelman (1999) (However,
see Nakamura 1998; Abramowicz, Lasota, \& Igumenshchev 2000). The 
Bernoulli parameter in our luminous hot solution is found to be in 
general negative. However, outflows are still possible in our model
when the vertical structure of the disk is taken into
account, or other factors such as magnetic field is included, 
as shown by the numerical simulation of Igumenshchev, Chen, \& 
Abramowicz (1996). 
In this case, we expect the presence of the transition 
from the outer new hot solution to the
inner ADAF in some situations. 

It was recently found that in addition to the self-similar ADAF solution
(Narayan \& Yi 1994), there is another self-similar CDAF (convection-dominated
accretion flow) solution which corresponds to a static envelope
in which the mass accretion rate is very small 
(Narayan, Igumenshchev, \& Abramowicz 2000;
Quataert \& Gruzinov 2000). This solution is possible only when 
the viscous parameter $\alpha$ and the direction or the efficiency of transport
of the angular momentum by convection satisfy some conditions (Narayan,
Igumenshchev, \& Abramowicz 2000). 
We don't know whether our luminous hot solution
is relevant to  CDAF since we don't know whether the luminous hot solutions
satisfy these two conditions. Almost all the analytical and numerical
simulations up to date are for ADAF (e.g., Narayan, Igumenshchev, \&
Abramowicz 2000, Quataert \& Gruzinov 2000; Igumenshchev, \& Abramowicz 1999;
Igumenshchev, Abramowicz, \& Narayan 2000). Our luminous hot
accretion solutions are physically quite different with ADAFs in the 
sense that the entropy of the flow in our solution decreases 
rather than increases towards the smaller radii.
This will have great impact on the convection 
stability of the flow. Therefore, 
the extension of study to the new
regime is needed.
  
\section{Promising Applications}
Obviously, this new hot solution is characterized by a high efficiency.
This together with its corresponding higher accretion rate compared to ADAF,
indicates that this type of hot accretion disk could emit a large amount
of hard X-ray radiation. This fact plus the arguments
given above about the formation channels of the luminous hot accretion solution,
make the solution a very promising model
for the X-ray luminous sources. An example we note is the 
very high state of black hole X-ray binaries.
Observations indicate that the luminosity of this state can reach 
close to Eddington luminosity, with the soft 
blackbody component and the hard nonthermal
power-law component being comparable in flux. The power-law component
does not show any evidence of a cutoff even out to a few hundred keV. 
The standard thin disk or the slim disk model 
can only explain the blackbody component.
ADAF, the only viable hot accretion disk model before the present paper,
can produce a hard power-law component, but its luminosity 
is too low compared with the blackbody component
because of its low accretion rate and efficiency (Esin, 
McClintock, \& Narayan 1997).
Our new hot accretion disk model with $\dot{m} \sim 1$ is a promising model.
The optically thick annulus formed within the transition radius
is responsible for the soft component in the spectrum of the very 
high state, while the hard component is
emitted by the hot corona above the annulus and the hot disk beyond the
transition radius. The high luminosity of the hard component is due to the
high efficiency and accretion rate while the extended power law hard tail is
because the electron energy distribution has a high energy power law 
tail as a result of the magnetic reconnection and shock acceleration in the
transition region. In addition, another signature of the very 
high state of BHXBs is their significant variability. 
The light curve seems to be composed of numerous random flares.
This is typically the feature of magnetic reconnection, which
must happens in the transition region as we argued in the previous section. 

We conjecture that the high state of BHXBs corresponds to the luminous
hot disk as well, but its accretion rate is smaller than in the very 
high state. The accretion rate may correspond to 
$\dot{m} \ga \dot{m}_2$ in this case. 
This will result in a moderately lower soft component luminosity and
much lower hard component luminosity compared to the
very high state. In addition, 
the luminous hot solution with 
$\dot{m}_1 \la \dot{m} \la \dot{m}_2$ might corresponds to 
certain low states in which the X-ray luminosity is 
moderately high, approaching 
$\sim 10\% L_{\rm Edd}$ (Nowak 1995). This is hard to be produced by an ADAF. 
Even though we assume $\alpha=0.3$, the highest luminosity of an ADAF 
is 0.1 ($ \approx \dot{m}_1$) $\times 10 \times 0.04$
($\approx$ efficiency) $\approx 4\% L_{\rm Edd}$. 

On the galactic scale, Seyfert 1 galaxies are in general assumed to be
the counterparts of the low state of BHXBs, and Narrow Line Seyfert 1
galaxies correspond to the very high state of BHXBs because of their
respective very similar spectra and variability features. 
It is thus an absorbing and also promising project to 
work out a unified satisfactory explanation 
to these sources by our luminous
hot accretion disk model. For example, similar with the very high state
of BHXBs, the Narrow Line Seyfert 1 galaxies could be interpreted by
our new hot accretion model with $\dot{m} \sim 1$. As shown by
Mineshige et al. (2000), the extreme soft X-ray excess can be 
accounted for by the slim disk model with $\dot{m} \sim 1$. 
The optically thick annulus in our model corresponds to 
the slim disk in Mineshige et al. (2000). While the hard X-ray
component whose flux is comparable to the soft X-ray component (Leighly 1999)
can be accounted for by the hot accretion flow outside 
the annulus and the hot corona above the annulus.  

\section*{Acknowledgments}
This work is supported by China Postdoctoral Fund and China 973 Project
NKBRSF G19990754.
I thank Mitch Begelman, Shin Mineshige,
 Ramesh Narayan, Eliot Quataert, 
Ronald Taam 
and Insu Yi for their helpful comments
and/or discussions. 
The hospitality of Korea Institute for
Advanced Study where this work began is acknowledged.

{}
\label{lastpage}
\end{document}